\documentclass{WileyMSP-template}
\usepackage{amssymb}
\usepackage{subfig}
\usepackage{svg}
\usepackage{amsmath}
\usepackage{float}
\usepackage{multicol}
\usepackage{comment}
\usepackage{lineno}
\usepackage{caption}
\usepackage{booktabs}
\usepackage{hyperref}

\begin{document}

%\pagestyle{fancy}
%\rhead{\includegraphics[width=2.5cm]{vch-logo.png}}

\title{Epitaxy of strained, nuclear-spin free $^{76}$Ge quantum wells from solid source materials}

\begin{center}
\maketitle

% Author: Please give full first and last names for authors and include * after the name of all corresponding authors
\author{
\\Maximilian Oezkent$^{1}$,
Chen-Hsun Lu$^{1}$,
Lucas Becker$^{2}$,
Sebastian Koelling$^{3}$,
Robert H. Blick$^{4}$,
Eloïse Rahier$^{3}$,
Stefan Schönert$^{5}$,
Nikolay Abrosimov$^{1}$,
Thilo Remmele$^{1}$,
Torsten Boeck$^{1}$,
Georg Schwalb$^{2}$,\\
Oussama Moutanabbir$^{3}$,
Martin Albrecht$^{1}$,
Carsten Richter$^{1}$,
Jens Martin$^{1}$,
Kevin-P. Gradwohl$^{1}$*
\\[1em]
$^{1}$Leibniz-Institut für Kristallzüchtung, Max-Born-Str. 2, 12489 Berlin, Germany \\
$^{2}$Siltronic AG, Johannes-Hess-Straße 24, 84489 Burghausen, Germany \\
$^{3}$Departmenct of Engineering Physics, École Polytechnique de Montréal, Succursale Centre-Ville, Montréal, H3C 3A7, Canada \\
$^{4}$Center for Hybrid Nanostructures (CHyN), Universität Hamburg, Luruper Chaussee 149, 22761 Hamburg, Germany \\
$^{5}$Physik Department, Technische Universität München, James-Franck-Straße 1, 85747 Garching, Germany
\\
*kevin-peter.gradwohl@ikz-berlin.de
}

% Keywords: Please provide a minimum of three and a maximum of seven keywords, separated by commas

\end{center}

% Abstract should be written in the present tense and impersonal style (i.e., avoid we), and be at most 200 words long
\begin{abstract}

Germanium quantum-well heterostructures have rapidly emerged as a leading platform for solid‑state quantum information processing; however, material quality limits scalability and higher structural quality, higher purity, as well as zero nuclear spin, are required. 
Here, we address these problems by employing the heaviest of Ge isotopes, by evaporating high-purity $^{76}$Ge radiation detector material, as utilized in fundamental neutrino particle physics experiments, to fabricate $^{76}$Ge/$^{28}$Si$^{76}$Ge quantum wells for quantum applications and explore the respective challenges. Specifically, we demonstrate improved results on strain-relaxed virtual Si$_{0.2}$Ge$_{0.8}$ substrates, forward graded from Si, with a dislocation density below $3.7 \cdot 10^{5}\,\mathrm{cm}^{-2}$, explore nuclear spin-free solid-source molecular beam epitaxy, and demonstrate first quantum transport in $^{76}$Ge quantum wells. We demonstrate a record-level quantum well interface width of $0.3~\mathrm{nm}$ by X-ray reflectivity, and quantitatively compare it to atom probe tomography and scanning transmission electron microscopy. The grown layer reveal nuclear-spin-bearing impurity concentrations below $10^{19}\,\mathrm{cm}^{-3}$ and chemical impurity levels below $10^{18}\,\mathrm{cm}^{-3}$, except for residual carbon attributed to the graphite crucible of the Ge source which may reach up to $10^{19}\,\mathrm{cm}^{-3}$. Low-temperature magneto-transport measurements yield electron mobilities of $6.1\cdot\mathrm{10}^4~\mathrm{cm^2V^{-1}s^{-1}}$ at 15~mK with a carrier density of $2.2\cdot10^{11}~\mathrm{cm^{-2}}$, indicating that residual carbon is the dominant scattering mechanism.

\end{abstract}

% Text: Please use section headings and subheadings as specified below. For communications, all section headings apart from Experimental Section should be removed
% Please make the first reference to a display item bold: \textbf{Figure 1}
% Do not abbreviate Figure, Equation, etc.; display items are always singular, i.e., Figure 1 and 2.
% Equations are always singular, i.e., Equation 1 and 2, and should be inserted using the {equation} environment, not as graphics
% Please do not use footnotes in the text, additional information can be added to the Reference list.

\section{Introduction}

In recent decades, group IV semiconductors silicon (Si) and germanium (Ge) have been crucial for the development of next-generation logic and memory devices \cite{pillarisettyAcademicIndustryResearch2011,goleyGermaniumBasedFieldEffect2014, hillerGroupIVSemiconductorMaterials2023}. While most electronic components remain Si-based, the importance of Ge layers and SiGe superlattice structures has steadily increased. For example, SiGe superlattices are essential for the fabrication of next generation C-FET devices \cite{schuddinckPPACSheetbasedCFET2022} and, due to their outstanding electrical and optical properties in conjunction with full compatibility to complementary metal–oxide–semiconductor (CMOS) technology. Furthermore Ge-based devices have advanced various fields including integrated photonic circuits \cite{marris-moriniGermaniumbasedIntegratedPhotonics2018, moutanabbirMonolithicInfraredSilicon2021} and high-mobility electronics \cite{toriumiGermaniumCMOSPotential2017}. High-purity Ge is also the material of choice for high-energy, high-resolution radiation detectors. Furthermore, such detectors made from the heaviest of Ge isotopes \textsuperscript{76}Ge, are employed in the most sophisticated neutrino experiments, answering fundamental questions in beyond the standard model physics \cite{abgrall2021legend}.

In recent years, Ge quantum wells (QW) have gained increasing attention due to their application in quantum technologies, specifically spin-based quantum computing \cite{scappucciGermaniumQuantumInformation2021}. Quantum bits (qubits) serve as basic building blocks whereby individual holes, i.e. non-occupied states in the Ge valence band, are confined in quantum dots (QDs). The QDs are defined vertically within a strained Ge QW and laterally by electrostatic gates. This device structure is characterized by a small spatial footprint, i.e. much smaller compared to superconducting qubits \cite{krasnokSuperconductingMicrowaveCavities2024} and offer the possibility for on-chip integration of control electronics \cite{ stehouwerExploitingEpitaxialStrained2024}. Paired with established cutting-edge CMOS fabrication, Ge-qubits promise a highly scalable solid-state quantum computing technology \cite{scappucciGermaniumQuantumInformation2021, stanoReviewPerformanceMetrics2022}. In this respect, an important step towards a scalable architecture, coherent shuttling of spin states along distances of several microns has been demonstrated \cite{vanriggelen-doelmanCoherentSpinQubit2024}. Overall, Ge-qubits have enabled outstanding progress due to their excellent quantum performance. Gate-defined hole spin qubits based on strained Ge quantum wells on relaxed SiGe offer several favorable properties. First, they exhibit inherently strong and tunable spin-orbit interaction, allowing for fast all-electrical qubit control, without, e.g. the use of micromagnets. The low effective mass of holes in Ge/SiGe leads to large quantum-dot level spacings, enabling larger device dimensions and thereby reducing lithographic fabrication requirements.\cite{hendrickxSingleholeSpinQubit2020, hendrickxFastTwoqubitLogic2020, hendrickxFourqubitGermaniumQuantum2021a, boganSingleHoleSpin2019, watzingerGermaniumHoleSpin2018, jirovecDynamicsHoleSingletTriplet2022, wangUltrafastCoherentControl2022, boscoFullyTunableHyperfine2021, delvecchioLightholeGatedefinedSpinorbit2023}

The suppressed hyperfine interaction arising from the p-orbital character of the valence band  of Ge naturally benefits long spin coherence of holes, but it can be further enhanced by isotopic purification to realize a nuclear spin-free host material, a critical requirement for quantum computing. 
Natural Ge and Si contain a fraction of non-zero nuclear spin isotopes, namely \textsuperscript{73}Ge and \textsuperscript{29}Si. Removing these nuclear spin fluctuations via isotope enrichment will enable improved sweet spot operation where qubit control is maximized while decoherence is minimized \cite{hendrickxSweetspotOperationGermanium2024a, stehouwerExploitingStrainedEpitaxial2025}. 

\subsection{The need for quantum-grade SiGe}
Hence, the need for nuclear spin-free material in spin qubits is a prime example to illustrate why standard electronic-grade SiGe is not sufficient for quantum applications. Classical electronic devices are not sensitive to nuclear spins; qubits are. Indeed, quantum applications also pose stricter boundaries on impurity levels, interface sharpness, and strain distributions. In order to highlight the need for improved material quality, and the complexity of parameters involved, we introduce the term "quantum-grade" as (yet) highest quality level of crystalline materials.

Achieving quantum-grade material quality remains challenging despite previous promising demonstrations on small qubit ensembles. Scale-up to the required large number of qubits demands further material and process improvements. 
For example, while some promising approaches of fabricating nuclear spin-free Ge/SiGe qubit host material via chemical vapor deposition (CVD) have been demonstrated recently \cite{moutanabbirNuclearSpinDepletedIsotopically2024a, daoustNuclearSpinfree70Ge2025}, the fabrication from solid source semiconductor sources requires further development. This applies also to pushing the limits of chemical and isotopic purity, atomically sharp interfaces, and structural perfection. 
Especially, interface sharpness directly influences strain distribution \cite{corley-wiciakNanoscaleMapping3D2023}, confinement potential, and carrier mobility—parameters crucial for maintaining high-fidelity spin control \cite{costaReducingDisorderGe2024}. Optimizing these aspects demands a precise understanding of the growth conditions and the substrate morphology, temperature-dependent interface formation, and possible sources of contamination.
Optimization of epitaxial processes must be tailored to the specific methodology. CVD and molecular beam epitaxy (MBE) are both well-established methods for SiGe fabrication \cite{tetznerDislocationsInfluenceBackground2025, beckerControllingRelaxationMechanism2020,nigroHighQualityGe2024,zhangHighqualityGeSiGe2024,eberlSiGeSi1994}. From a fabrication perspective, CVD is an industrially scalable method capable of producing high-quality epitaxial layers. However, achieving isotopic purity with CVD remains challenging due to the limited availability of isotope-enriched precursor gases \cite{ernstSilicon28TetrafluorideEductIsotopeEngineered2024}. Furthermore, the growth of strained Ge quantum wells using CVD can be unfavorable because the required elevated temperatures promote strain relaxation, leading to dislocation formation. In contrast, MBE, particularly solid-source MBE, enables the growth of isotopically enriched layers with a high compositional and thickness control and a wide growth temperature window, making it well-suited for the development of quantum-grade material. 

\subsection{The motivation for \textsuperscript{76}Ge MBE}
The central approach of this work, is to employ the highly developed, isotope-enriched, high-purity \textsuperscript{76}Ge crystals developed within the LEGEND collaboration, and employ them in our cutting-edge MBE setup to produce the next generation quantum-grade SiGe, and the first ever demonstration of \textsuperscript{76}Ge-based quantum materials. The influence on nuclear spin of Si is avoided by using \textsuperscript{28}Si solid source material. The details on the source material specifications are found in the Supporting Information S1.

We employ a hybrid approach that leverages the scalability of CVD for substrate preparation and the precision of solid-source MBE for the growth of isotope-purified active layers, which represents a promising and also efficient process to obtain quantum-grade heterostructures with a reduces usage of costly enriched sources. We start the discussion by presenting improved results on CVD-grown strain-relaxed buffer (SRB) virtual substrates, and continue with the first SiGe heterostructures achieved with an even-isotope-only Si/Ge-MBE system. Interface sharpness and surface roughness as a function of growth temperature are investigated within the constraint of achieving fully strained Ge QW layers on CVD-grown Si$_{0.2}$Ge$_{0.8}$ SRB, considering the impact of the substrate quality. Magneto-transport measurements are utilized to investigate scattering mechanisms of charge carriers in the nuclear spin-free heterostructures by molecular beam epitaxy. Through this approach, we demonstrate pathways toward a material platform that meets the stringent requirements for next-generation quantum devices based on Ge hole spins. 

On a completely different note, one has to mention that such \textsuperscript{76}Ge-based materials are relevant due to the high isotopic mass, regarding Ge-phononics and the impact of isotope mass on superconducting transition temperature of highly p-doped Ge.

\section{Results and Discussion}

\subsection{CVD of SRB}

\begin{figure}[H]
    \centering
    \includegraphics[scale = 1]{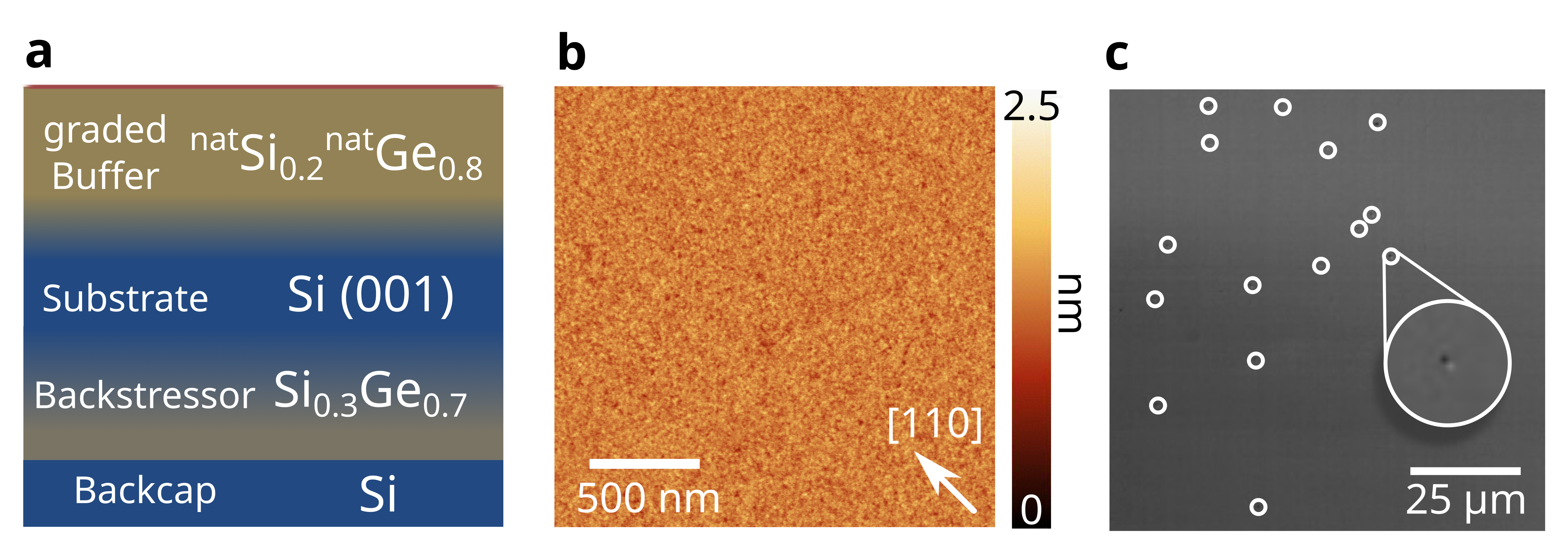}
    \caption{CVD grown virtual substrate properties with (a) a cross-section schematic of the layer stack with the Si substrate, the backstessor and the graded Buffer. (b) An AFM image scanned $\approx$ 45° to [1\,1\,0\,] of the epi-ready and cleaned surface with an $\sigma_{rms} =$ 0.15~nm. (c) A SECCO etched sample micrograph from the center of a 12" SRB with TDs marked white and a closeup on a TD. On a etched 500 x 500 µm$^2$ area the TD are counted resulting in a TDD = 3.7$\cdot10^{5}~cm^{-2}$.}
    \label{fig:SRB}
\end{figure}

A schematic of the SRB layer stack can be seen in \textbf{Figure \ref{fig:SRB}a}. As the Ge concentration increases, the lattice parameter in plane and out of plane increases accordingly and eventually plateaus within the fully relaxed buffer layer. The intermediate layers of constant composition, namely: Si$_{0.94}$Ge$_{0.06}$, Si$_{0.7}$Ge$_{0.3}$ Si$_{0.52}$Ge$_{0.48}$ , Si$_{0.34}$Ge$_{0.66}$, and finished by Si$_{0.2}$Ge$_{0.8}$ are highlighted. Those steps are close to the Si radial cut, which implies a proportional increase of lattice parameters, which in turn implies a relaxation. From the constant composition Si$_{0.2}$Ge$_{0.8}$ peak, a lattice parameter in and out of plane are determined to be 5.60~Å. Which is in good agreement with the lattice parameter calculated by a parabolic interpolation \cite{dismukesLatticeParamLeterDensity}. The according RSM in a (2~2~4) $Q_{\perp}$ vs $Q_\parallel$ plot can bee seen in \textbf{Figure \ref{fig:HS}d}. After applying the wet-chemical cleaning of the CMP-polished, epi-ready SRB surface, the surface appears randomly rough. This can be seen in an AFM image of a representative area of the sample in Figure \ref{fig:SRB}b. No cross-hatch pattern is distinguishable in this and also in larger scans, which are not shown here. The resulting roughness is $\sigma_{RMS}=0.15$~nm. Figure \ref{fig:SRB}c is a DIC micrograph after a Secco etch with a white circle indicating the TDD.  In such SRBs Becker et al. \cite{beckerControllingRelaxationMechanism2020} pointed out that the dislocation-driven relaxation starts at preferential nucleation sites on the edge of the wafer. By adjusting the growth parameters, the gliding velocity of the dislocation is controlled, and in the case of a Si$_{0.2}$Ge$_{0.8}$, it is achieved to get a TDD of $3.7\cdot10^{5}~cm^{-2}$ in the center of the wafer.

\subsection{Optimized Ge quantum well epitaxy}
The nuclear spin-free Ge QW heterostructures were developed by MBE on these SiGe SRBs. One central challenges is achieving well-defined QW interfaces. Hereby, the compressively strained Ge quantum well tends to facet and roughen if the epitaxy temperature is sufficiently high, which is driven by the consequent partial release of elastic energy. This can be influenced by the choice of the growth temperature $T_{g}$. This is systematically investigated here and optimized to achieve the lowest interface width possible.

\begin{figure}[H]
    \centering
    \includegraphics[width = 1 \linewidth]{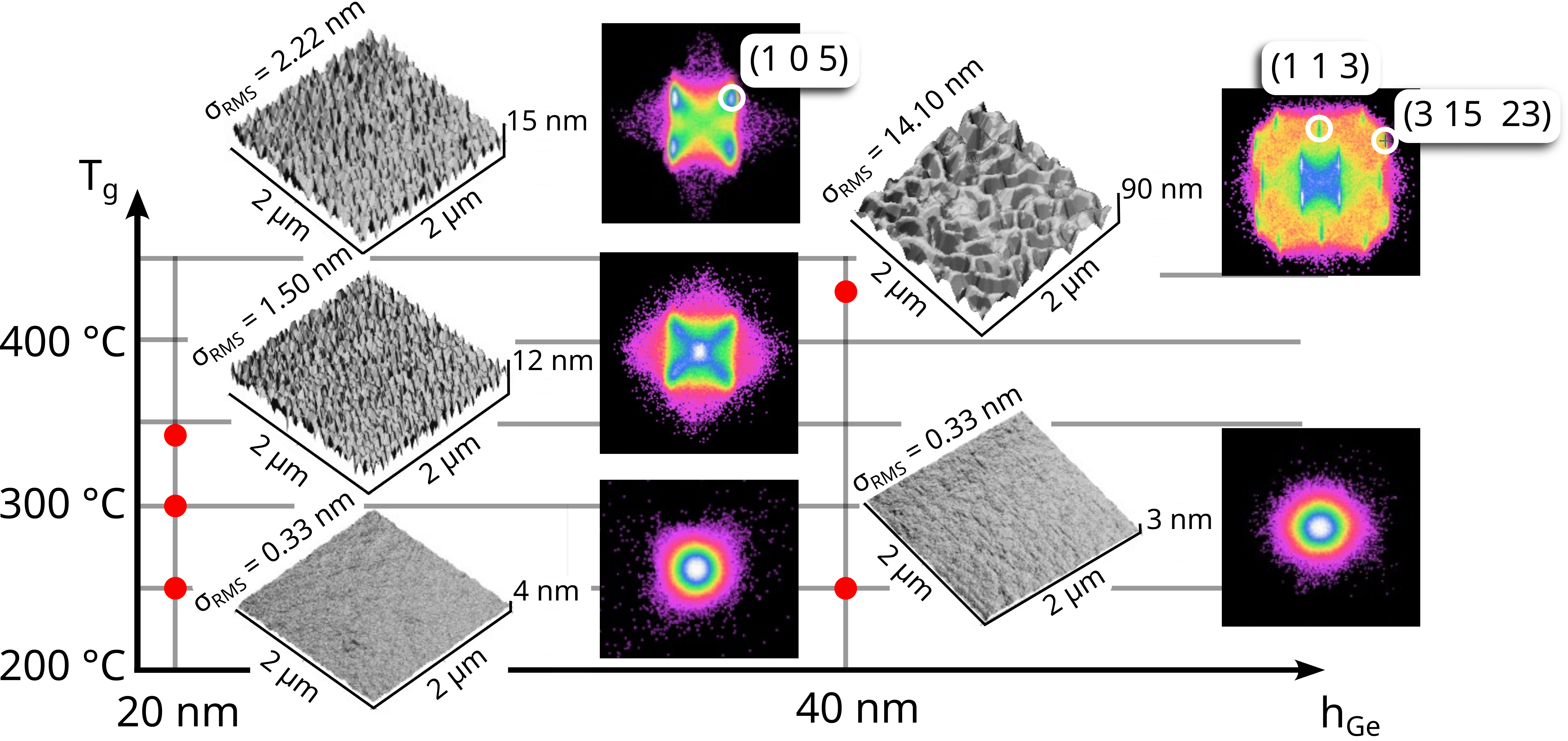}%
    \caption{AFM based morphology analyses of strained Ge layer grown on Si$_{0.3}$Ge$_{0.7}$ SRB as a function of layer thickness and temperature. Thereby facet analyses shows $\{$1 0 5$\}$, $\{$1 1 3$\}$ and $\{$3 15 23$\}$ facets which are indexed accordingly.}
\label{fig:HS_growth_window}
\end{figure}

For this, a Ge QW growth series on Si$_{0.3}$Ge$_{0.7}$ SRB with varying $T_{g}$ at a growth rate of 0.02~nm/s is performed. The results are shown in Figure \ref{fig:HS_growth_window}. The transfer of the results to the 80~\% SRB is valid since the lattice parameter differs only slightly (5.59~$\text{\AA}$ vs. 5.60~$\text{\AA}$) and the SRB is fabricated the same way, which results in the same quality.

Growing h$_{Ge}$ = 20~nm Ge at 250~°C results in a surface with an $\sigma_{RMS} = 0.33$~nm and no faceting can be determined. This also holds by doubling the layer thickness to 40~nm. Increasing the growth temperature to 300~°C, surface rippling can be determined in view of the 3D morphology which is reflected in the increasing surface roughness $\sigma_{RMS} = 1.50$~nm. The facet analysis shows first signs of $\{$1 0 5$\}$ facets appearing. This gets more pronounced in an growth at 340~°C. There the facets analysis reveals pronounced $\{$1 0 5$\}$ facets with in an surface roughness of$\sigma_{RMS} = 2.22$~nm. In order to explore the faceting behavior, 40~nm is grown at 430~°C. This layer exhibits a complete ridge-and-valley morphology where the faces a faceted with $\{$1 0 5$\}$, $\{$1 1 3$\}$ and $\{$3 15 23$\}$ facets and a surface roughness of $\sigma_{RMS} = 2.22$~nm.
The ridge‑and‑valley morphology is attributed to strain relaxation in lattice‑mismatched epitaxial growth. In this case, it becomes energetically favorable for the surface to roughen by forming ridges, despite the associated increase in surface area, in order to relieve the mismatch‑induced strain. This leads to a transition in the growth regime toward island growth with pronounced facets. This island growth regime becomes more pronounced with increasing temperature, resulting in enhanced surface rippling and facet formation.\cite{chenStructuralTransitionLargeLatticeMismatch1996, rastelliSurfaceEvolutionFaceted2002}

Based on those results, a full Ge QW heterostructure temperature ramp is carried out which is shown in the Figure \ref{fig:HS}b. This ramp leads, at a growth rate of 0.02~nm/s, to the best structural results without getting kinetically induced roughening or faceting.
The whole structure is grown without growth interrupt or shutter movement, and ramping $T_{g}$ is done during the growth process. This prevents a shutter movement induced burst of matter (peak at 2$\cdot$$10^{-10}$~mbar) and also the formation of further growth interfaces (absorption of impurity species during growth interrupt). Lowering the temperature at the BB/QW interface allows for the growth of a fully-strained Ge QWs. In order to suppress Ge-segregation into the TB, the temperature is kept low until several monolayers of SiGe are grown \cite{gradwohlEnhancedNanoscaleGe2025}. 

%Capping the whole layer with $<$~2~nm of highly strained Si \cite{wyssBulksuppressedSurfacesensitiveRaman2024} is done on either room temperature (RT), 90~°C, 240~°C, or 320~°C with different outcomes on crystallinity, see Section \ref{sub:Si-cap}.

\subsection{Ge QW characterization}
\begin{figure}[H]
    \centering
    \includegraphics[width = 1 \linewidth]{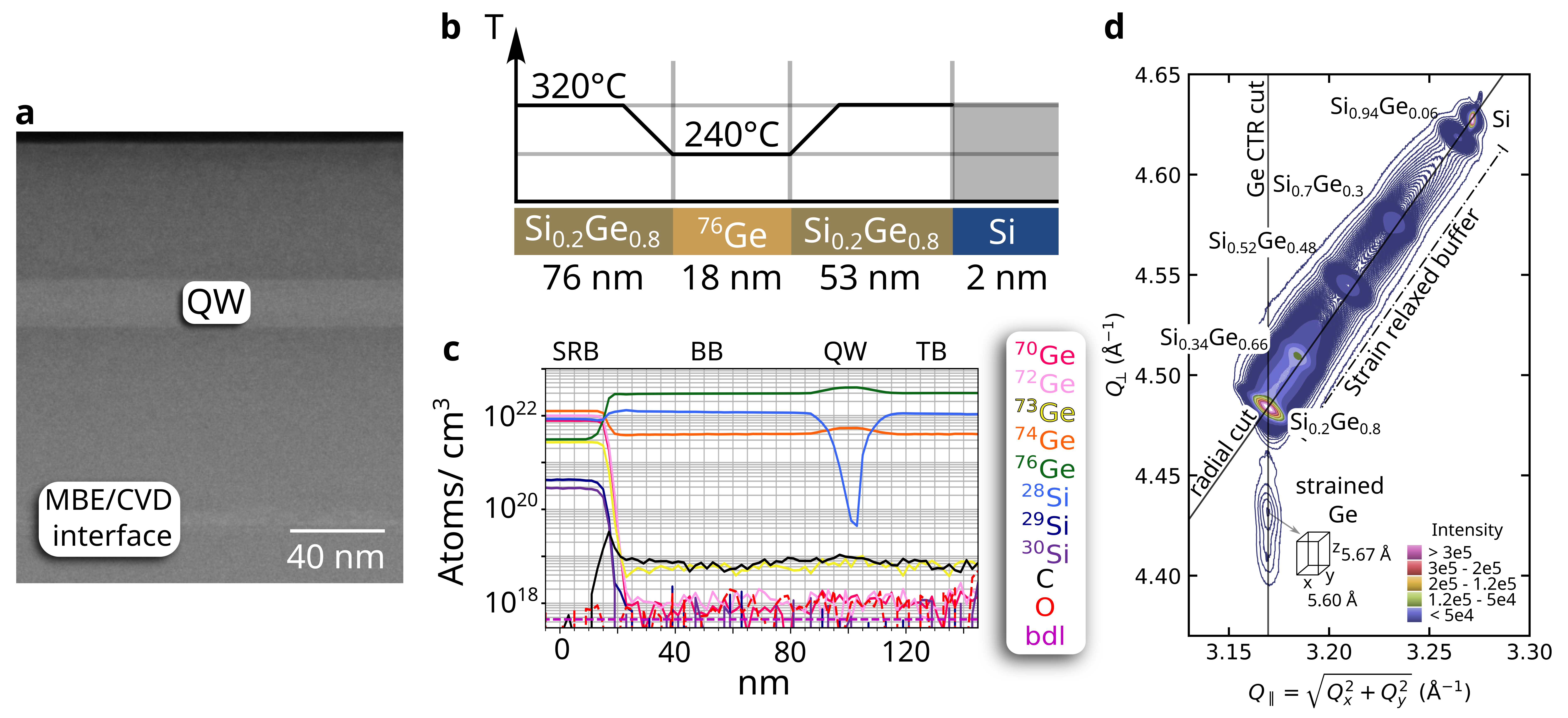}
    \caption{Structural characterization of a nuclear-spin-free Ge/SiGe heterostructure. (a) Cross-sectional STEM image of the heterostructure, and the according growth temperature profile during the MBE process with the possible to change the Si-cap appearance by changing the growth temperature in (b). (c) APT mass spectrometry data showing the distribution of relevant isotopic masses along the heterostructure. In (b) a reciprocal space map (RSM) in (2~2~4) reflection is displayed as $Q_{\perp}$ vs. Q$_{\parallel}$. The strained Ge QW is located along the Ge CTR cut. The calculated lattice parameters are shown.}
\label{fig:HS}
\end{figure}

The optimized Ge QW heterostructures on these SRBs are displayed in a cross-sectional STEM image in Figure \ref{fig:HS}a. Based on the image, no larger structural inhomogeneities in the heterostructure can be attested. The CVD/MBE interface indicates a Ge-rich interface layer, which is likely explained by the selective etching of Si during the cleaning procedure. There is no evidence that further growth is affected by this interface feature. Figure \ref{fig:HS}b shows the applied growth temperature profile, which leads to the shown heterostructure.

The APT results are shown as a profile in Figure \ref{fig:HS}c, having the SRB with the abundant isotope proportion of Si and Ge, C and O below detection limit (bdl) on the right. A detailed discussion of impurity incorporation into the grown layer under the present conditions and taking into account the concept of impingement rate, is provided in the Supporting Information S2. In view of the silicon isotope ratio, $^{30}$Si and $^{29}$Si concentration drop below the detection limit in the MBE-grown layers. The interface width is broad, and the $^{28}$Si is still present in the QWL since the data is binned for noise reduction reasons. This has the drawback of less depth-dependent compositional precision determination. The Ge isotope ratio changes from the SRB into the MBE grown layer. Isotopes $^{70,72,73}$Ge are each several magnitudes less in concentration $<$~10$^{19}$~Atoms/cm$^3$ in the MBE layer versus in the SRB $>$~10$^{21}$~Atoms/cm$^3$.
The tensile strain state of the Ge QW is confirmed by a (2~2~4) RSM seen in \ref{fig:HS}d. Along the Ge crystal truncation rod (CTR) cut $\parallel$ to $Q_{\perp}$, the signal of the Ge QW is located, which indicates a good crystalline quality without any signs of relaxation. The compressively strained Ge lattice parameter is increased according to the Poisson effect in the out-of-plane axis $z$ of $\sim$ 1.2\% (5.60~$\text{\AA}$ vs. 5.67~$\text{\AA}$).

\begin{figure}[H]
    \centering
    \includegraphics[width = 1 \linewidth]{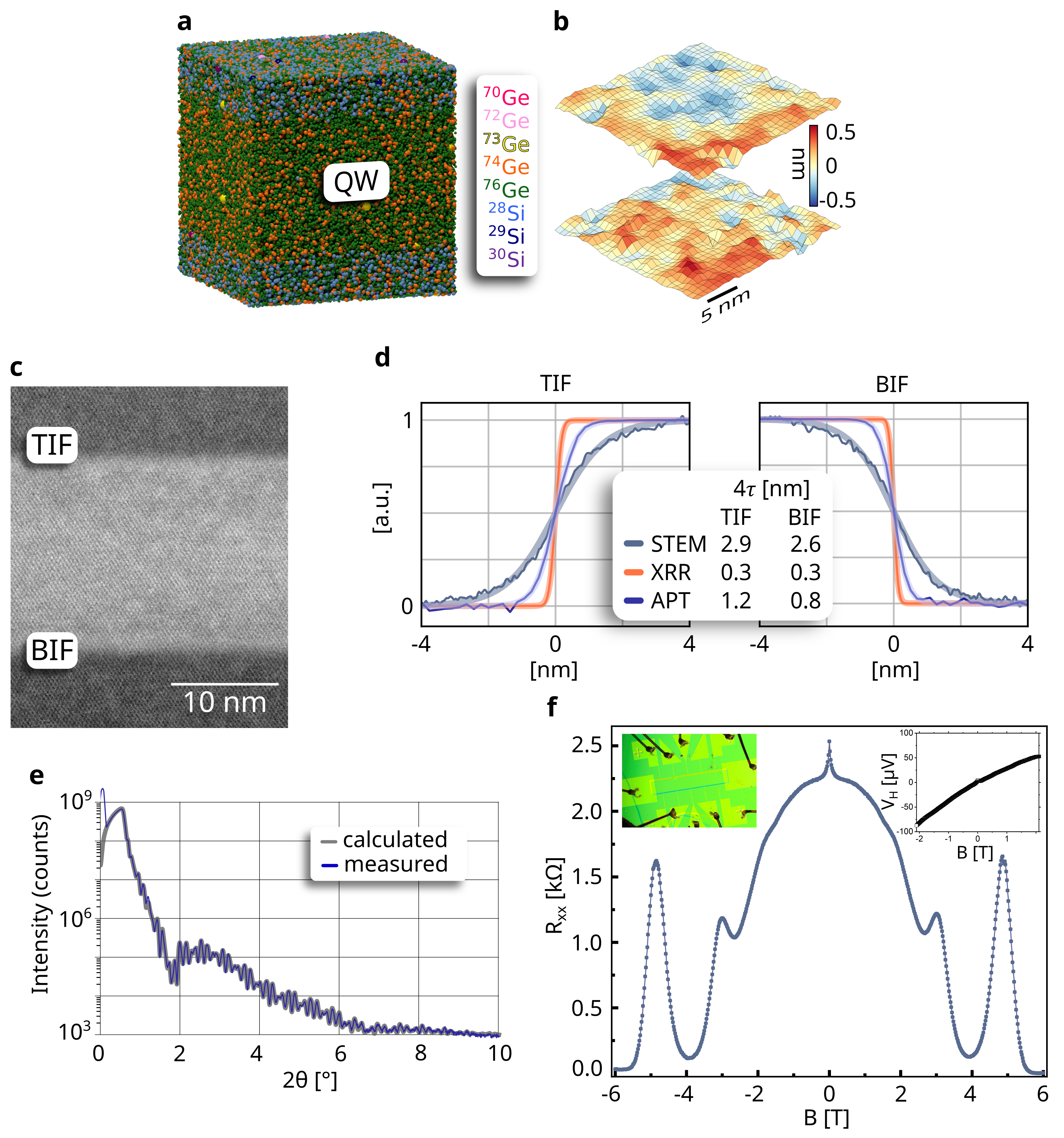}
    \caption{ 
    (a) Rendered model from 3D APT data where each point denotes the estimated position of an ionized atom detected.
    (b) represents isosurfaces with the BIF at the bottom and TIF at the top. Ge concentrations for the isosurface are BIF at Ge: 87.6\% and TIF at Ge: 87.7\%. This isosurface spans a 30~x~30$~nm^2$ area, which represents the lateral size of a quantum dot, and the corresponding $\sigma_{RMS}$ values are BIF: 0.154~nm and TIF: 0.154~nm.
    (c) STEM image with a closeup of the Ge QW and marked TIF and BIF and a simulated ideal interface SIF.
    (d) Comparison of interface sharpness for the TIF and BIF at 4$\tau$, derived from STEM, XRR, and APT analyses, presented as a length versus normalized value plot. The lighter (less opaque) lines indicate the corresponding sigmoid function fits.
    (e) XRR measurement data and the corresponding calculated fit displayed in an intensity vs. 2$\theta$ plot ($\sigma_{RMS}$ values for BIF: 0.106~nm, TIF: 0.147~nm).
    (f) Low-temperature magneto-transport measurement on a Ge/SiGe 2DEG at 15 mK and zero back-gate voltage. With image insert of the Hall bar geometry on the top left and the Hall slope plot at the top right.}
\label{fig:Interface}
\end{figure}

The quantification of a Ge QW interfaces comes with major challenges since the desired interface width is in the regime of $<$ 1~nm. Approaches to determine such interface widths are performed on Si-QWs and Ge QWs via APT and STEM or both \cite{penaModelingSiSiGe2024, degliespostiLowDisorderHigh2024, koellingThreeDimensionalAtomicScaleTomography2023, paqueletwuetzAtomicFluctuationsLifting2022c, dyckAccurateQuantificationSi2017a}. Our approach integrates STEM, APT, and XRR to characterize the interface, with the goal of quantitatively assessing interface sharpness with an 4$\tau$ value and using these insights to optimize growth parameters. \textbf{Figure \ref{fig:Interface}} shows the results of those methods.
In order to get an intuitive and qualitative way of seeing the Ge QW with its isotope distribution, a 3D rendered model is shown. Thereby Figure \ref{fig:Interface}a depicts a 3D point-cloud reconstruction of the Ge QW based on APT data, in which each point denotes the estimated position of an ionized atom detected.

A QW interface width determination is performed on a concentration profile along the growth direction from a reconstructed 3D APT dataset.\cite{basGeneralProtocolReconstruction1995, paqueletwuetzAtomicFluctuationsLifting2022b}. An iso-surface is created at the Ge \% value at which a sigmoidal fit along the profile has its inflection point. This is shown in Figure \ref{fig:Interface}b where the bottom surface represents the bottom interface (BIF) at Ge: 87.6\% and the top surface the top interface (TIF) at Ge: 87.7\%. This isosurface spans a 30~x~30~$nm^2$ area which represent approximately the size of a quantum dot. \cite{paqueletwuetzAtomicFluctuationsLifting2022} 
The isosurfaces indicate no larger fluctuations and appear homogeneous rough.
In consideration of a close-up STEM image which is shown in Figure \ref{fig:Interface}c, the analysis of the QW interface is also performed by a sigmoid error function fit along the growth direction.
Further, an investigation with XRR is performed, which allows for broader statistics over the illuminated area (in $cm^2$ range). The resulting measured profile along 2$\theta$ is shown in Figure \ref{fig:Interface}e including the fitted curve. The gained information includes for each layer in the MBE grown structure the density, thickness and interface roughness. The density and therefore SiGe composition and also the thickness information are consistent with the gained STEM and APT values.
The comparison of the interface shape gained by APT, STEM and XRR with the according 4$\tau$ values are shown in Figure \ref{fig:Interface}d. The 4$\tau$ values for each method are gained from the fitted sigmoid curve (lighter color). The gained data show a divergence in appearance (interface broadness) between the methods which is also indicated by the 4$\tau$ values. The STEM based interface analyses appear the most broad, which is mainly due to noise arising from the lamella thickness (>~200~nm). The interface measured by APT appears narrower, but factors arising during data acquisition and subsequent analysis, such as positional uncertainty and detection efficiency, must be taken into account. The sharpest apparent interface is observed with XRR because the technique probes a large lateral area and measures the averaged electron density profile perpendicular to the surface. Small-scale atomic fluctuations that influence APT measurements play a comparatively minor role in XRR, resulting in a smoother and sharper measured interface. Other factors such as surface roughness or interfacial mixing beyond the instrument resolution can still broaden the interface in XRR, but under typical conditions, XRR tends to yield the sharpest apparent interface for the same sample.
Those methods should be treated complementary when it comes to evaluating the interface and a absolute interface roughness value in this regime cannot be given.

\subsection{Si capping layer}
\label{sub:Si-cap}

The purpose of the Si cap is to create a well-defined and controlled surface for subsequent device fabrication and suppress unintentional charge noise by reducing interface trap formation. Such traps arises from inhomogeneous Ge and Si sub-oxide formation therefore it requires a Ge-free Si-cap.

Consequently, the challenges of Si-cap epitaxy on 80\% Ge buffers are manifold. The over 3\% strain limit the layer thickness of relaxation-free caps to 2~nm. Ge segregation from the SiGe TB into the cap makes it difficult to incorporate the necessary SiGe concentration gradients from 80\% to essentially 0\% Ge in these few monolayers of cap layer. Here, we systematically investigate epitaxy of $^{28}$Si caps in $^{28}$Si$^{76}$Ge barriers.

\begin{figure}[H]
    \centering
    \includegraphics[width = 1 \linewidth]{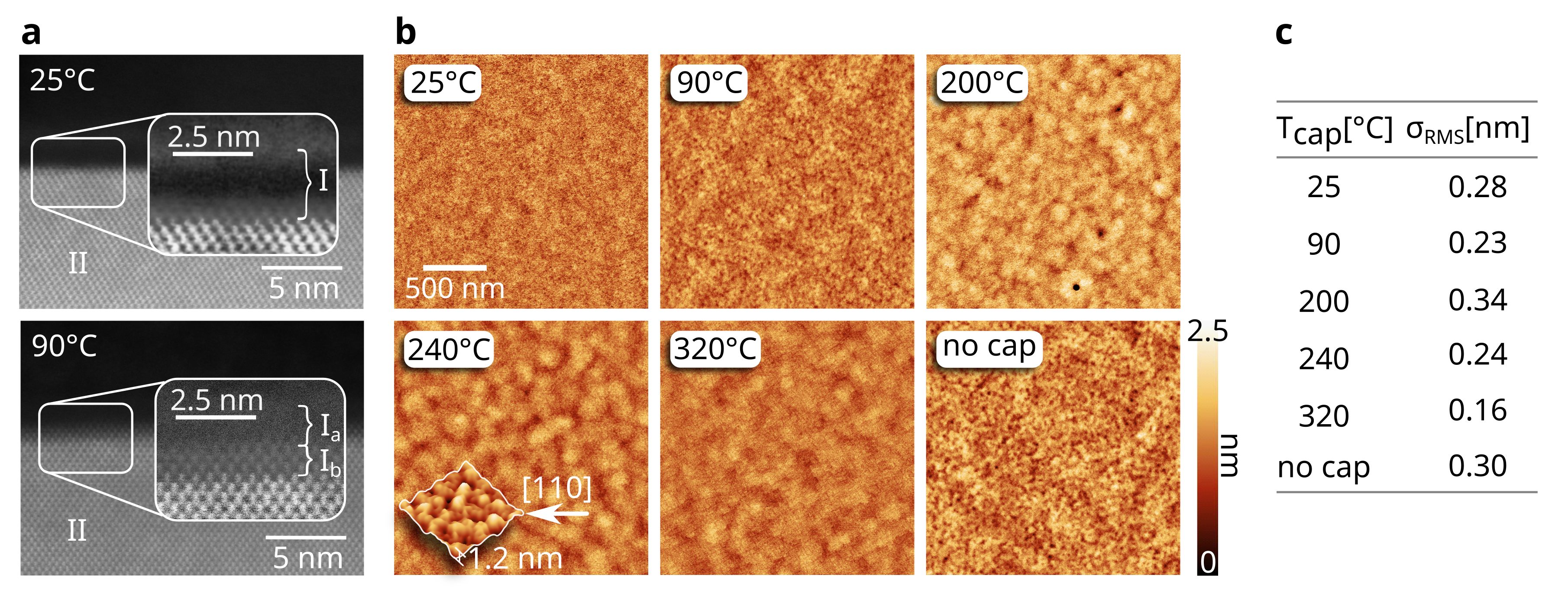}%
    \caption{(a) Si-cap surfaces grown under different growth temperatures shown in close up STEM images of the Si-cap for an amorphous cap grown at 25~°C and a crystalline, partially oxidized Si-cap grown at 90~°C. I, I$_a$ and I$_b$ show the different Si-cap appearances in a closed up STEM image an II the underlying top barrier. (b) 2$\cdot$2~µm$^2$ AFM images with the according growth Temperature. At 240~°C a 3D representation of the surface morphology is shown. The according $\sigma_{RMS}$ for each AFM meausrement are shown in the table in (c).}
    \label{fig:Si-cap}
\end{figure}

In Figure \ref{fig:Si-cap}b, AFM images of a 2~nm Si-cap temperature series grown on top of a Ge QW heterostructure are shown. The AFM measurements were taken immediately after growth, with an ambient-air exposure time of less than 30~min. For reference, an uncapped heterostructure is also shown. The resulting $\sigma_{RMS}$~values of each temperature step is shown in Figure \ref{fig:Si-cap}c. 
A $T_{cap}$ of 25~°C results in a randomly rough surface with $\sigma_{RMS} = 0.28$~nm. The growth interruption required to bridge the temperature gap between 320~°C TB growth and Si-cap deposition exceeds 10~h cool-down time and involves shutter movement. At this temperature a amorphous cap is assumed where a STEM cross-sectional image of the sample surface shown in Figure \ref{fig:Si-cap} supporting this assumption. There is no contrast indicating in the Si$_{0.2}$Ge$_{0.8}$ layer at the bottom (II) a crystalline Si layer. By focusing on the top region and applying contrast-limited adaptive histogram equalization to enhance local contrast, a faint interface contrast appears at a position consistent with the expected 2~nm Si-cap height I. This contrast originates from Fresnel fringes based on the interface between amorphous Si and the glue layer. The exact width, shape, and position of this interface depend, among other factors, on beam defocus and are therefore only of qualitative relevance.
In contrast, the STEM analysis of the $T_{cap}$~=~90~°C sample indicates distinct contrast variations in the top regions, $I_a$ and $I_b$. While $I_a$ exhibits contrast comparable to that of region $I$, $I_b$ appears notably brighter than the surrounding TB region, suggesting the presence of a crystalline Si-cap beneath an amorphous and partially oxidized Si.
At $T_{cap}$~=~240°C underlying surface undulations get the most pronounced with a maximum amplitude of approximately 1.2~nm which is highlighted as a three-dimensional surface in the AFM image. Such rounded features may originate from either the onset of strain-induced rippling (cf. Figure \ref{fig:HS_growth_window}) or from Ge segregation into the cap layer, which results in a fluctuating surface morphology.
Growing the cap at 320~°C results in the formation of terraces ($\sigma_{RMS} = 0.16$~nm), as the enhanced surface diffusion enables Si adatoms to occupy energetically favorable sites but also increases the Ge-segregation into the cap.
In general, Si-caps exhibit smoother surfaces than uncapped structures due to the highly strained nature of the thin Si layer. Since a capping temperature of $T_\text{cap}$~=~240~°C does not require a growth interrupt and yields a smooth, non‑terraced cap, future device structures will be capped under these conditions.

\subsection{Magneto-transport measurements}

Low-temperature magneto-transport measurements performed on the Ge/SiGe two-dimensional electron gas (2DEG) at \( T = 15 \, \text{mK} \) and zero back-gate voltage reveal a smooth, symmetric longitudinal resistance \( R(B) \) with a pronounced maximum at zero magnetic field which is shown in Figure \ref{fig:Interface}h. This zero-field peak is characteristic of weak localization and reflects phase-coherent backscattering in a diffusive transport regime. The magnetoresistance evolves monotonically with increasing \( |B| \), with signatures of Shubnikov–de Haas oscillations in the upper field range, indicating moderate carrier density and mobility.
For the measurements we applied an AC excitation voltage of \( 1 \, \text{V} \) over a \( 100 \, \text{M}\Omega \) pre-resistor, yielding a current of \( 10 \, \text{nA} \). The Hall signal amplitude was about \( 50 \, \mu\text{V} \). The Hall bar geometry is \( W = 20 \, \mu\text{m} \), \( L = 100 \, \mu\text{m} \) (\( W/L = 0.2 \)) and is shown in the left insert of the Figure. Longitudinal resistance at zero field was about \( R_{xx} \approx 2.35 \, \text{k}\Omega \). From the Hall slope \( dV_{xy}/dB = 2.85 \times 10^{-5} \, \text{V/T} \), corresponding to \( dR_{xy}/dB = 2.85 \times 10^{3} \, \Omega/\text{T} \). This yields a carrier density of \( n_s \approx 2.2 \times 10^{11} \, \text{cm}^{-2} \). Based on the sheet resistivity of \( \rho \approx 470 \, \Omega/cm \), we obtain a mobility of \( \mu \approx 6.1 \times 10^4 \, \text{cm}^2/\text{Vs} \).
Those results, especially the carrier density with the measured carbon concentration, let assume that residual carbon is the dominant scattering mechanism in the Ge QW.

\section{Conclusion}
We demonstrated epitaxy of nuclear spin-free $^{76}$Ge/$^{28}$Si$^{76}$Ge quantum well heterostructures using solid source material in MBE with growth pressure below $4$ $\cdot$ 10$^{-10}$~mbar, and illuminated the related growth challenges and consequent material quality, such as interface constitution, chemical and isotopic purity, and magneto-transport properties. 
The SRBs SiGe buffers grown by have, determined by etch pit density investigation, a TDD of 3.7 $\cdot$ 10$^5~$cm$^{-2}$ at a nominal composition of Si$_{0.2}$Ge$_{0.8}$. Later was determined by reciprocal space maps. 
This substrate enabling a high-quality nuclear spin-free MBE without using isotope-enriched material besides the rather thin active $^{76}$Ge quantum well heterostructures. 
A MBE growth window was developed which takes the Ge segregation, surface roughness, and the background impurity contamination into account. This results in fully strained $^{76}$Ge QW heterostuctures with a sharp top ($4\tau_{_{TIF}}: 0.3 - 2.9~nm$) and bottom ($4\tau_{_{TIF}}: 0.3 - 2.6~nm$) interface. Furthermore, epitaxy of optimized $^{28}$Si-cap layer on Si$_{0.2}$Ge$_{0.8}$ are investigated. A quantitative assessment and comparison of the sharpness was carried out based on the complimentary methods XRR, APT and STEM.
Isotopic- and chemical purity were investigated by APT, and showed nuclear spin concentrations below 10$^{19}$~cm$^{-3}$, and concentration of all impurities except carbon below detection limit of 10$^{18}$~cm$^{-3}$. The residual carbon concentration can be attributed to the graphite crucible of the Ge source, which was concluded by analyzing the absorption rate of carbon during epitaxy as determined by in-situ mass spectroscopy. 
Additionally, low-temperature magneto-transport transport measurements were conducted and indicate a mobility of $\mu~\approx~6.1~\cdot~\mathrm{10}^4~\mathrm{cm^2/Vs}$ at 15~mK with a carrier density of $n_s~\approx~2.2~\cdot~10^{11}$, which is attributed to scattering of charge carriers on the residual C concentration.
% Experimental section

\section{Experimental Section}

\subsection{CVD}
We employ CVD-grown SiGe SRB fabricated using chlorinated precursors, which enable the highest-quality SRBs to be achieved at high growth rates and elevated growth temperatures. The CVD grown, 12" SRB is schematically displayed in \textbf{Figure \ref{fig:SRB}a}, with Si$(0\,0\,1)$ as a substrate, the Si$_{0.3}$Ge$_{0.7}$ bottom layer as a back-stressor including a Si backcap, the linear graded buffer with progressively increasing Ge content and the Si$_{0.2}$Ge$_{0.8}$ top layer of constant composition which is in total serval hundred $\mathrm{\mu}$m thick \cite{beckerThreadingDislocationDensity2024,beckerControllingRelaxationMechanism2020}. The SRBs were grown in a CVD reactor (ASM E3200) on 300 mm Si wafers using the precursors GeCl$_4$, SiH$_2$Cl$_2$ and H$_2$ as the carrier gas. The deposition took place at high temperatures ($>$800~°C) and atmospheric pressure conditions. Prior to epitaxy, the wafers underwent a chemo-mechanical polishing step in order to smooth out the pronounced cross-hatch pattern.

\subsection{MBE}
A detailed discussion of the key components required for the fabrication of nuclear-spin-free Ge/SiGe heterostructures using nuclear spin-free solid source $^{76}$Ge and $^{28}$Si can be found in the Supporting Information S1 and S3.

The utilized MBE chamber has a base pressure of $3\cdot10^{-11}$ mbar and the residual gases are analyzed by a quadrupole mass spectrometer (HiQuad Quadrupole Mass Spectrometer QMG 700) with a detection limit around $10^{-14}$ mbar and a detection range up to 100~u. The mass spectrometer is placed underneath the sample at a distance of 50~cm and tilted at a slight angle, enabling it to detect emitted species from the sample surface efficiently. Using a liquid nitrogen (LN$_2$) cryo-shroud brings the chamber walls to 77 K, resulting that only 2~u is still detectable in idle. Then, a growth pressure of $4\cdot10^{-10}$ mbar is achieved with detected masses of 2~u (molecular hydrogen) being the dominant species by orders of magnitude, followed by 28~u as a secondary detectable gas species in the $^{28}$Si MBE setup. 

\begin{figure}[H]
    \centering
    \includegraphics[width = 1 \linewidth]{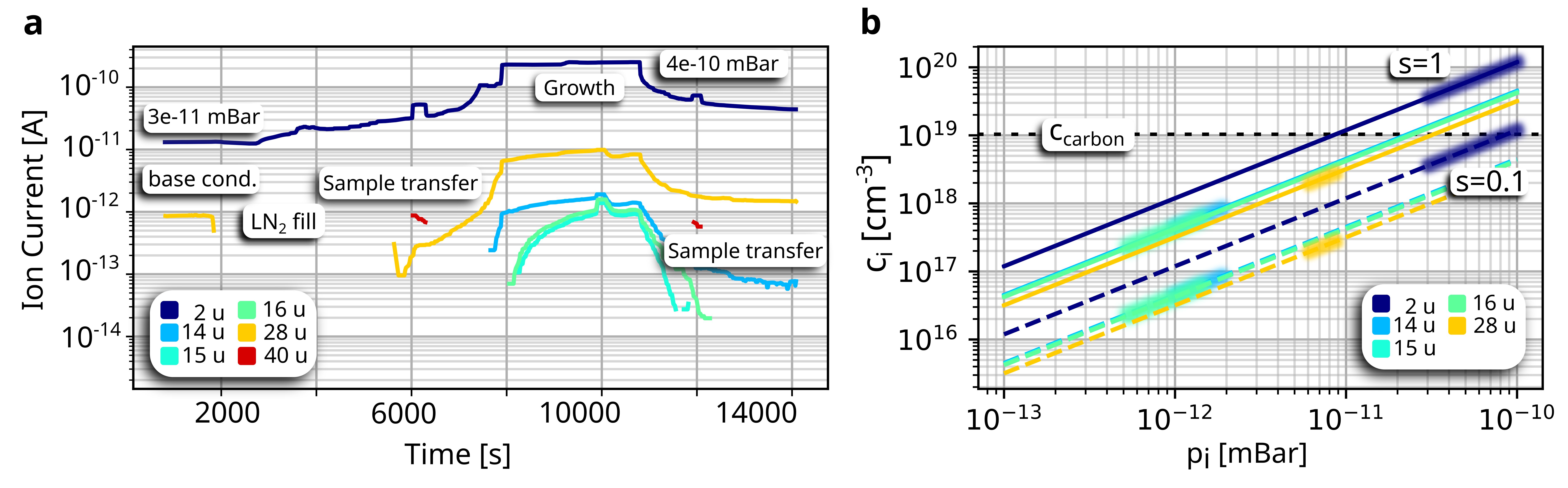}
    \caption{(a) Mass spectrometer read out from base state with a base pressure of 3e-11 mbar. Continuing cooling down the MBE walls with LN$_2$ fill, up-to the growth initiation and the growth end. (b) Calculated impurity concentration as a function of partial pressure for different molecular masses. The curves correspond to measured masses in (a), with two surface sticking coefficients s~=~1 (solid lines) and s~=~0.1 (dashed lines).}
    \label{fig:Impurity}
\end{figure}

The 12" SRB wafers were cleaved and saw-cut to 25~x~25~mm$^2$ sample pieces. Prior to growth, the substrates underwent wet-chemical cleaning. Starting with a degreasing step in an ultrasonic bath with Acetone and n-Propanol, subsequently. Followed by a 60~s dip in a hydrofluoric acid (HF 50\%) 1:4 solution in order to remove residual organic contamination and native oxides. The step also hydride-passivates the surface, which is noticeable by the hydrophobic behavior. \cite{sunSurfaceTerminationRoughness2006,ponathGe001Surface2017,kagawaChemicalEtchingGermanium1982,digaspareEpicleaningGeGeSn2015,ganScanningTunnelingMicroscopy1998} After each step, a 15~M$\Omega$cm deionized water rinse is performed. All used solutions meet an ultra-high purity electronic-grade standard. The exposure time to ambient air between the last cleaning step and the loading it into a glovebox load lock is $<$~15~s. The glovebox has a 5N Argon atmosphere in which the substrate is transferred to the UHV load lock carrier. This step minimizes the ambient atmospheric contamination and utilizes the cleaner UHV conditions. The substrate is annealed to 650~°C for 10~min in a preparation chamber ($10^{-10}$~mbar) in order to desorb hydrogen, residual oxygen, and possible carbon species from the substrate surface.

\subsection{Characterization methods}
\textit{AFM}\\
The morphology of the grown layers is investigated by a Bruker Dimension Icon atomic force microscope (AFM) in peak force tapping mode. The evaluation and data processing of the AFM data is performed with the software Gwyddion \cite{necasGwyddionOpensourceSoftware2012} including a data correction to reduce measurement artifacts by applying mean plane subtraction and a third-degree polynomial data correction. Additionally 3D data displays are chosen to highlight surface morphologies in a qualitative matter.
Facet analysis was performed using the Gwyddion Facet Analysis tool, which determines the distribution of local surface orientations by calculating surface normals with respect to the mean surface, and grouping them into facets according to their slope and direction (see Gwyddion manual for details). The angle information of the facets are used to determine and indexing the {h k l} of the facets.
The scans are performed $\approx$ 45° to [1\,1\,0] to avoid misinterpretation of cross hatch pattern and line by line variations of the AFM scan.
\\\textit{XRD}\\
In order to disclose the chemical, isotopic and structural properties of the grown heterostructures, several characterization techniques are applied. X-ray diffraction (XRD) is used to investigate strain of the Ge QW by utilizing a reciprocal space map (RSM) at the $(2\,2\,4\,)$ reflex. X-ray reflectometry (XRR) and subsequently data modeling is used to determine layer thickness, interface roughness and indicate layer composition. XRR and XRD were conducted in Rigaku's SmartLab x-ray diffractometer, equipped with a Cu K$_\alpha$ anode, operating at a voltage and current of 45~kV and 200~mA, and a HyPix-3000 hybrid pixel array detector. For XRR measurements, the instrument was configured in the parallel beam geometry utilizing soller slits of 0.25° and a receiving slit of 0.2~mm as a trade-off between angular resolution and signal strength. For XRD experiments, the instrument was operated in high-resolution mode without receiving optics to maximize the captured diffraction signal. The XRR data were modeled using the pyXRR python package \cite{richterCarichtePyxrrFirst2017}, which utilizes Fresnel equations with Parratt-Formalism for simulating reflectivity, and lmfit, which employs least-squares minimization approach for curve fitting.
\\\textit{STEM}\\
Scanning transmission electron microscopy (STEM) is used to determine the thickness and composition of the layers as well as the width of the interfaces. The STEM investigations were carried out using an aberration corrected FEI Titan 80-300 TEM, operated at 300~kV. The angular range for the ADF detector collection was nominally 36-150~mrad. Atomic-resolution ADF-STEM images were captured along the $[1~1~0]$ zone axis. Sample preparation was performed via mechanical polishing and conventional ion milling in Gatan’s Precision Ion Polishing System (PIPS) to achieve electron transparency.
Atom-probe tomography (APT) is used to determine not only the layer thickness, layer composition, interface roughness but also depth dependent isotope composition. 
\\\textit{APT}\\
The APT samples were prepared in a FEI Helios Nanolab 600 dual-beam scanning electron microscope using a gallium-focused ion beam at 30~kV. The samples were cooled down to a temperature of 25~K and where measured in an CAMECA Invizo 6000.
\\\textit{Defect Etching}\\
A 2~min selective defect etching after Secco \cite{seccodaragonaDislocationEtch1001972} in combination with an microscope in differential interference contrast (DIC) mode is used to reveal threading dislocation density (TDD) by counting them in a 500~x~500~$\mu m^2$ representative area.

\medskip
\textbf{Supporting Information} \par %Please delete the Suppporting Information statement if it is not applicable. Please supply Supporting Information in another file. Supporting information should not be provided in .tex format
Supporting Information is available from the Wiley Online Library or from the author.
\medskip

\textbf{Acknowledgements} \par
We acknowledge the financial support provided by the European COST Action OPERA (CA-20116) through the short-term scientific missions program and the financial support by the Federal Ministry of Education and Research (BMBF) of Germany in the project QUASAR (Project No. FKZ 13N15659). Furthermore we acknowledge the financial support by the European research council (ERC) via a ERC Advanced Grant GemX \cite{ERC_GemX}.
\medskip
\\\textbf{Conflict of interest}\par
The authors declare no conflict of interest.
\medskip
\\\textbf{Data Availability Statement}\par
The data that support the findings of this study are available from the corresponding author upon reasonable request.

% References
\medskip

% Use the following code if you wish to generate your bibliography with BibTeX;
% replace the string "MSP-template" below with the name(s) of
% the BibTeX data base(s) you want to use.
% The resulting bibliography-output (the content of the .bbl file)
% must be pasted back into this file before submission.
% Please also include your BibTeX data base file(s) in your submission
% so that we can re-run BibTeX if necessary.
%
\bibliographystyle{MSP}
\bibliography{References}
\bigskip
%\textbf{Figures}\\

\medskip
\begin{figure}[H]
    \centering
    \textbf{Table of Contents}\\
    \medskip
    \includegraphics[scale = 1]{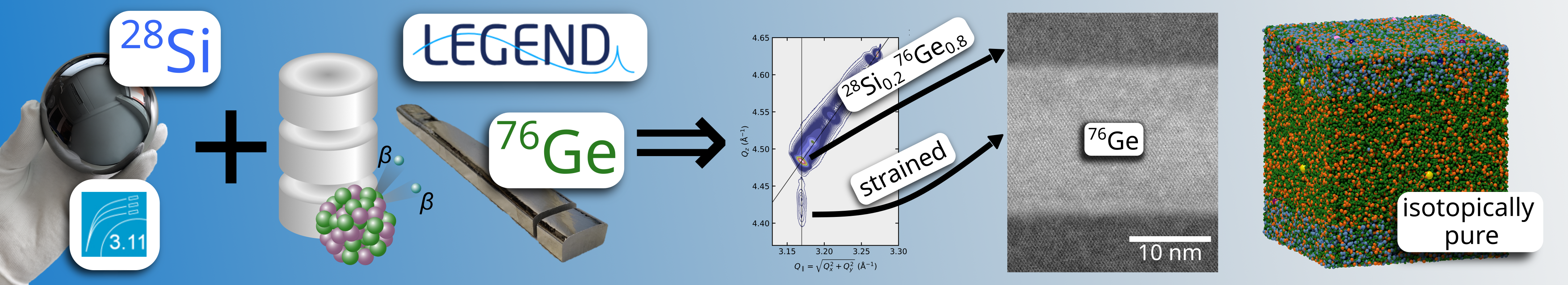}
    \medskip
    \caption{Spin-based quantum computing still suffers from limited coherence times, often constrained by nuclear spin noise. Isotopically purified materials offer a pathway forward. By utilizing ultra-pure $^{76}$Ge from the LEGEND experiment and $^{28}$Si from the Avogadro project, we establish a quantum-grade host platform. With optimized fabrication, this approach enables a new generation of scalable, low-noise quantum materials.}
\end{figure}

% Figures/tables and captions
% Permission statements are required for all figures reproduced or adapted from previously published articles/sources. Please also ensure that all necessary permissions to reproduce images have been received
% Please remove these statements for original figures

% Table of contents entry should be 50 - 60 words long
% Image should be 55 mm broad and 50 mm high or 110 mm broad and 20 mm high

\end{document}